\documentstyle[multicol,prl,aps,epsfig]{revtex}

\def\sumdash{\mathop{\sum{}^{'}}}
\def\proddash{\mathop{\prod{}^{'}}}
\newcommand{\bsigma}{\mbox{\boldmath $\sigma$}}
\renewcommand{\narrowtext}{\begin{multicols}{2}
\global\columnwidth20.5pc\noindent}
\renewcommand{\widetext}{\end{multicols}\global\columnwidth42.5pc}
\begin{document}
\draft
\title{
Vertical Density Matrix Algorithm: A Higher-Dimensional Numerical
Renormalization Scheme based on the Tensor Product State Ansatz
}
\author{
Nobuya Maeshima, Yasuhiro Hieida${}^1$, Yasuhiro Akutsu, Tomotoshi
Nishino${}^1$, and Kouichi Okunishi${}^2$
}
\address{
Department of Physics, Graduate School of Science, Osaka University, Toyonaka,
560-0043, Japan\\ 
${}^1$Department of Physics, Graduate School of Science, Kobe University,
Rokkoudai, 657-8501, Japan\\ 
${}^2$Department of Physics, Faculty of Science, Niigata University, Igarashi,
950-2181, Japan\\ 
}
\date{\today}
\maketitle
\begin{abstract}
We present a new algorithm to calculate the thermodynamic quantities of
three-dimensional (3D) classical statistical systems, based on the ideas of the
tensor product state and the density matrix renormalization group. 
We represent the maximum-eigenvalue eigenstate of the transfer matrix as the
product of local tensors which are iteratively optimized by the use of the
``vertical density matrix'' formed by cutting the system along the transfer
direction.  This algorithm, which we call {\em vertical density matrix
algorithm} (VDMA), is successfully applied to the 3D Ising model.  Using the
Suzuki-Trotter transformation, we can also apply the VDMA to two-dimensional
(2D) quantum systems, which we demonstrate for the 2D transverse field Ising
model. 
\end{abstract}


\section{Introduction}
Since the density matrix renormalization group (DMRG) method was invented by
S.R.White, ~\cite{orig-DMRG1} the method has been applied to various problems in
one-dimensional (1D) quantum systems and two-dimensional (2D) classical
systems.~\cite{dmrg-rev} Such a great success of the DMRG has been stimulating
us to extend the algorithm to the one which can handle higher-dimensional
systems, principally 2D quantum systems and 3D classical
systems.~\cite{LiPa,Nishino-Okunishi-CTTRG,Henelius,Martin} 

We should recall that, in the DMRG, the matrix-product structure of the
wavefunction of the target states (usually the ground state or the
maximum-eigenvalue eigenstate) is essential.~\cite{Ostlund}  From this point of
view, the tensor product state (TPS) which is a natural higher-dimensional
generalization of the matrix product state, should play a key role in the
``higher-dimensional DMRG''.  A simple but non-trivial example of the TPS is the
ground state of the 2D valence-bond-solid (VBS) type quantum spin systems, where
the wavefunction is expressed as a product of local finite-dimensional tensors,
with all the tensor indices being contracted.~\cite{tasaki}  As for 3D classical
statistical systems, the maximum-eigenvalue eigenstate of the layer-to-layer
transfer matrix of the 3D classical system can exactly be represented as the TPS,
if we allow the tensor dimension to be infinite.  We should note that we can
reduce the calculation of the expectation value of the TPS to a statistical
average in a lower-dimensional classical system;  $(D+1)$-dimensional classical
(or $D$-dimensional quantum) problem reduces to a $D$-dimensional classical
statistical problem.\cite{njp} In fact, the properties of the 2D VBS model have
been studied in terms of a 2D vertex model associated with the TPS
wavefunction.~\cite{njp,Niggemann}

When developing the TPS formulation, the most important step is the
determination (optimization) of the local tensor in the TPS.  In
Refs.~\cite{Okunishi-Nishino-KW,SCTPVA}, direct variational formulations are
employed for the optimization of the local tensors.  As compared with these
``direct'' method, our novel algorithm given in the present paper for 3D
classical system is more like the original DMRG.  The local tensor is updated
by the ``block-spin basis transformation'' along the vertical direction.  Since
this transformation is constructed in terms of the density matrix made along the
``vertical direction'' (transfer direction of the transfer matrix), we call this
algorithm the {\em vertical density matrix algorithm} (VDMA).  We apply the VDMA
to the 3D Ising model and  discuss its efficiency.  We also report the
application of the VDMA to the 2D transverse field Ising model, with help of the
Suzuki-Trotter transformation~\cite{Suzuki}.

This paper is organized as follows: In Sec.\ref{sec:explain-concept}, we briefly
explain the VDMA for 3D classical spin systems, taking the 3D Ising model as an
example.  In Sec.\ref{sec:result-3dIsing} we show the numerical result for the
3D Ising model and the 2D transverse field Ising model. The last section is
devoted to the conclusion.

\section{Method}
\label{sec:explain-concept}
\begin{figure}
\epsfxsize=2.6 in\centerline{\epsffile{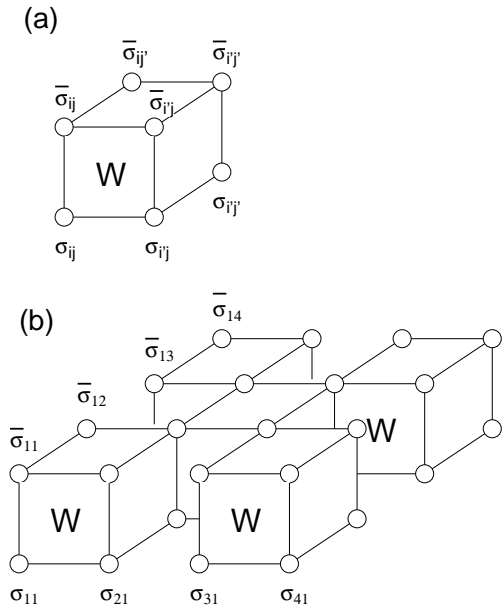}}
\caption{(a) The Boltzmann weight of a unit cell, and (b) the transfer matrix 
$T[\bar{\sigma}|\sigma]$ ($N=4$).}
\label{fig:transmat}
\end{figure}

Let us consider the 3D Ising model on the simple cubic lattice of the size
$N \times N \times 2L$ in  X, Y, and Z directions. Suppose that $L$ and $N$ are
sufficiently large, and the neighboring Ising spins $\sigma$ and $\sigma'$ have
ferromagnetic interaction $-J \sigma \sigma'$. Then the Boltzmann weight for the
unit cube is written as 
\begin{eqnarray}
W\left( \begin{array}{cccc}
\bar{\sigma}_{ij}&\bar{\sigma}_{i'j}&\bar{\sigma}_{i'j'}&\bar{\sigma}_{ij'} \\
\sigma_{ij}&\sigma_{i'j}&\sigma_{i'j'}&\sigma_{ij'} \\ \end{array} \right)
  \equiv  \exp \biggl[ -\frac{K}{2}(\sigma_{ij}\sigma_{i'j} 
 +\sigma_{i'j}\sigma_{i'j'} 
+\sigma_{i'j'}\sigma_{ij'}+\sigma_{ij'}\sigma_{ij}+
\bar{\sigma}_{ij}\bar{\sigma}_{i'j}
+\bar{\sigma}_{i'j}\bar{\sigma}_{i'j'} \nonumber \\
+\bar{\sigma}_{i'j'}\bar{\sigma}_{ij'}+\bar{\sigma}_{ij'}\bar{\sigma}_{ij}
+\sigma_{ij}\bar{\sigma}_{ij}
+\sigma_{i'j}\bar{\sigma}_{i'j}+\sigma_{i'j'}\bar{\sigma}_{i'j'}
+\sigma_{ij'}\bar{\sigma}_{ij'}  ) \biggr],\label{eq:defBW}
\end{eqnarray}
where  $i'=i+1$, $j'=j+1$, and $K=J/T$.
The locations of the spin variables are shown in Fig.~\ref{fig:transmat}.

For the book keeping, let us here introduce some notations for spin
variables. Write the configuration of the four spins surrounding a plaquette in
the $XY$ plane as 
\begin{equation}
 \{\sigma_{ij}\} = (\sigma_{ij} \sigma_{i'j} \sigma_{i'j'} \sigma_{ij'} ),
\end{equation}
where the position of the plaquette can be labeled by the index $ij$. Then the
Boltzmann weight is  simply written  as 
$W \left\{ \matrix{ \bar{\sigma}_{ij} \cr  \sigma_{ij} \cr} \right\}$.
Also for a spin layer in the $XY$ plane, we denote the configuration of the
$N\times N$ spins as, 
\begin{equation}
[ \sigma ] \equiv \left[
\begin{array}{ccc}
\sigma_{11} & \ldots  & \sigma_{1N} \\
\vdots      & \ddots  &  \vdots     \\
\sigma_{N1} & \ldots  & \sigma_{NN} \\
\end{array}
\right].
\label{lspin}
\end{equation}
Using these notations, the transfer matrix $T$ from a layer $[\sigma]$ to the
next layer $[\bar{\sigma}]$ is written as:
\begin{equation}
T[\bar{\sigma}|\sigma] =\prod_{ i+j=even}
W \left\{ \matrix{ \bar{\sigma}_{ij} \cr \sigma_{ij} \cr} \right\}  \label{eq:deftm}.
\end{equation}
In the product of Eq.~(\ref{eq:deftm}),  the spin variables are shared by the
adjacent cubes in the diagonal direction and thus the Boltzmann weights make the
checkerboard pattern in the $XY$-plane.(see Fig.1-(b))


Our goal is to evaluate the maximum-eigenvalue $\lambda_{\rm max}$ of the
transfer matrix $T$ and the corresponding eigenvector $|\psi_{\rm max}\rangle$,
using the TPS representation of the eigenvector. In order to analyze the TPS
structure of $|\psi_{\rm max}\rangle$,  let us  consider the power method
briefly, which is the simplest but powerful technique to calculate
$\lambda_{\rm max}$ and $|\psi_{\rm max}\rangle$.
Define the vector $|\psi_L\rangle$ with
\begin{equation}
 |\psi_{L}\rangle = T^L|\psi_{0}\rangle,
\label{eq:power}
\end{equation}
where $|\psi_{0}\rangle$ is an `initial' vector that is not orthogonal to
$|\psi_{\rm max}\rangle$.
Then the maximum-eigenvalue eigenvector $|\psi_{\rm max}\rangle$ is obtained as
\begin{equation}
|\psi_{\rm max} \rangle =  {\rm const} \times  \lim_{L \to \infty} |\psi_{L}
 \rangle,
\end{equation}
where the details of  $|\psi_{0}\rangle$ are not important.

\begin{figure}
\epsfxsize=2.6 in\centerline{\epsffile{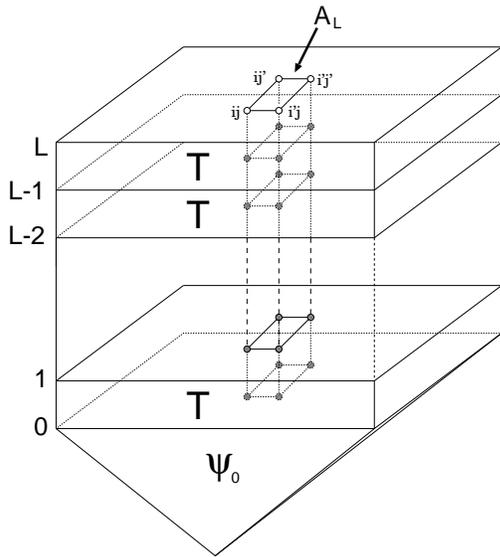}}
\caption{The graphical representation of the vector $|\psi_L\rangle$. Below the
$ij$-plaquette, we  find that the  $L$ unit cubes are piling up.}
\label{fig:power}
\end{figure}

In Fig.~\ref{fig:power}, we show the  graphical representation of
$|\psi_{L}\rangle$, where we can see the structure of $|\psi_{L}\rangle$ more
clearly. As is seen in this figure, the $L$ unit cubes are piling vertically up
to the surface, and then the product of the vertically-arranged Boltzmann
weights can be regarded as a local tensor; We define the local tensor  $A_L$ at
the $ij$-plaquette as,
\begin{equation}
A_L\left\{ \matrix{ \sigma^L_{ij} \cr \xi^L_{ij} \cr } \right\} = 
W\left\{ \matrix{ \sigma^L_{ij}  \cr  \sigma_{ij}^{L-1} } \right\}
W\left\{ \matrix{ \sigma_{ij}^{L-1} \cr \sigma_{ij}^{L-2} } \right\}
\cdots
W\left\{ \matrix{ \sigma_{ij}^{1} \cr \sigma_{ij}^0 } \right\},
\end{equation}
where  $\{\sigma_{ij}^L\}$ is the spin configuration  at the  $ij$-plaquette on
the ``surface layer'', and the auxiliary variable 
$\xi^L_{ij}\equiv(\sigma_{ij}^0 \sigma_{ij}^1 \sigma_{ij}^2 \cdots
\sigma_{ij}^{L-1})$ denotes the configuration for the spins under the surface.
Using the local tensors defined above,  $|\psi_L \rangle$ is represented as a TPS:
\begin{equation}
|\psi_L \rangle=
\sum_{[\xi_L]} \prod_{i+j=even} A_L\left\{ \matrix{ \sigma^L_{ij}  \cr
\xi^L_{ij} \cr } \right\}.
\end{equation}
Taking the limit $L \rightarrow \infty$, we have the maximum-eigenvalue
eigenvector $|\psi_{\rm max} \rangle$, which is now  represented as the product
of the local tensor $A_{\infty}$. 


From the practical view point of the usual power method,  the eigenvector is
improved  with the relation $|\psi_{L+1}\rangle=T|\psi_{L}\rangle $ iteratively.
In the framework of the TPS, this power method is reformulated as the recursion
relation for the local tensor:
\begin{equation}
A_{L+1} \left\{ \matrix{ \sigma^{L+1}_{ij} \cr \xi^{L+1}_{ij} } \right\} =
W \left\{ \matrix{ \sigma^{L+1}_{ij} \cr \sigma^{L}_{ij} } \right\}
A_{L} \left\{ \matrix{ \sigma^{L}_{ij} \cr \xi^{L}_{ij} } \right\},
\label{eq:extension}
\end{equation}
with which we can carry out the iterative calculation, until $A_L$ gives a good
approximation of $A_\infty$. However it is generally difficult to store the
tensor $A_L$ in the computer memory for a  sufficiently large $L$, because if
$\xi^{L}_{ij}$ has $M$ states, then the extended  auxiliary variable $\xi^{L+1}$
has $2M$ states.

\begin{figure}
\epsfxsize=2.6 in\centerline{\epsffile{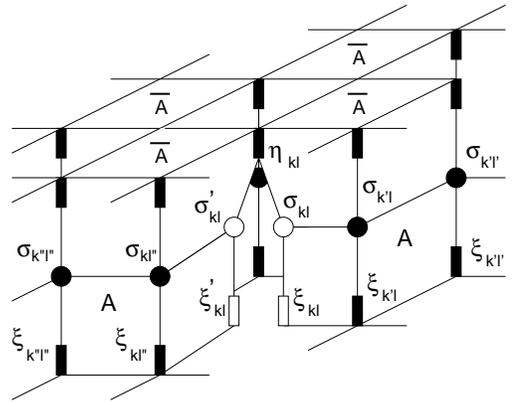}}
\caption{The vertical density matrix $\rho_{kl}(\sigma'_{kl}
\xi'_{kl}|\sigma_{kl} \xi_{kl})$.} 
\label{fig:vdm}
\end{figure}

In order to restrict the number of the auxiliary variable, we now import the
idea of the DMRG into the TPS.\cite{orig-DMRG1} The essence of the DMRG is that
the increased  number of states for $\xi^{L+1}_{ij}$ can be reduced, by using
the ``projection operator'' generated from the ``density matrix''. 
In the present case,  the appropriate density matrix should be constructed for
the spin variables $(\sigma^{L+1}_{ij}, \xi^{L+1}_{ij})$ in the vertical
direction (we thus call this density matrix as ``the vertical density
matrix'').\cite{BdotB} Introducing the ``transposed'' local tensor:
\begin{equation}
\bar{A} \left\{ \matrix{\eta_{ij} \cr \sigma_{ij} \cr} \right\} 
 = W\left\{ \matrix{ \bar{\sigma}^0_{ij} \cr \bar{\sigma}^{1}_{ij} \cr} \right\} 
   W\left\{ \matrix{ \bar{\sigma}^{1}_{ij} \cr \bar{\sigma}^{2}_{ij} \cr}
 \right\}   \cdots
   W\left\{ \matrix{ \bar{\sigma}^{L-1}_{ij} \cr \sigma_{ij} \cr} \right\},
\end{equation}
the explicit form of the vertical density matrix is defined as (See
Fig~\ref{fig:vdm}.),
\begin{eqnarray}
\rho_{kl}(\sigma'_{kl} \xi'_{kl}|\sigma_{kl} \xi_{kl})&=&
\sumdash_{[\sigma],[\xi],[\eta]}
\left[\proddash_{i+j=even}
\bar{A} \left\{ \matrix{\eta_{ij}\cr \sigma_{ij}\cr }\right\}
{A} \left\{ \matrix{ \sigma_{ij} \cr \xi_{ij}\cr } \right\} \right]
  \bar{A}\left\{ \matrix{ \eta_{k''l''}\cr \check{\sigma}_{k''l''}\cr }\right\}
       A \left\{ \matrix{ \check{\sigma}_{k''l''} \cr \check{\xi}_{k''l''}\cr }
\right\} \nonumber \\
&\times&  \bar{A}\left\{ \matrix{ \eta_{kl}\cr \sigma_{kl}\cr }\right\}
 A \left\{ \matrix{ \sigma_{kl} \cr \xi_{kl}\cr } \right\} \label{eq:getrho} ,
\end{eqnarray}
where $\sum^{'}$ denotes  the configuration sum for the all spin variables
except  $\sigma_{kl}^{(')},\xi_{kl}^{(')}$, and  $\prod^{'}$ means the product
for the site index $(i,j)$  except at $(i,j)=  (k,l),\; (k'',l'')$ with
$k''=k-1$ and $l''=l-1$. In Eq. (\ref{eq:getrho}), we have also used the
notation for the ``checked spins'': 
\[ \left\{ \matrix{ \check{\sigma}_{k''l''} \cr \check{\xi}_{k''l''}\cr }
\right\} \equiv\left( \begin{array}{cccc}
\sigma_{k''l''} & \sigma_{kl''} & \sigma'_{kl} & \sigma_{k''l} \\
\xi_{k''l''} & \xi_{kl''} & \xi'_{kl} & \xi_{k''l} \end{array} \right).
\]
Further, we have omitted the sub(super)script $L$ if it is apparent.


For the case of the isotropic 3D Ising model, the vertical density matrix
$\rho_{kl}$ becomes independent of the site index $kl$ in the thermodynamic
limit. Thus we write  $\rho_{kl}$ simply  as $\rho$  hereafter. Moreover it
should be noted that, for the isotropic case, the local tensors $A$ and
$\bar{A}$  satisfies the relation: 
\begin{equation}
 \bar{A}\left( \begin{array}{cccc}\eta_{ij} & \eta_{i'j} & \eta_{i'j'} & \eta_{ij'} \\
\sigma_{ij} & \sigma_{i'j} & \sigma_{i'j'} & \sigma_{ij'} \end{array} \right)= 
 A\left( \begin{array}{cccc} \sigma_{ij} & \sigma_{ij'} & \sigma_{i'j'} & \sigma_{i'j} \\
\eta_{ij} & \eta_{ij'} & \eta_{i'j'} & \eta_{i'j} \end{array} \right).
\end{equation}

Following the spirit of the DMRG, we diagonalize $\rho$ to have the eigenvalues
$w_{\tilde{\xi}}$ in the decreasing order $w_1 \ge w_2 \ge \cdots (\ge 0)$:
\begin{equation}
 \sum_{\sigma \xi} \rho (\sigma' \xi'|\sigma \xi) U( \sigma \xi|\tilde{\xi})
= U(\sigma' \xi'|\tilde{\xi} )  w_{\tilde{\xi}},
\end{equation}
where  $U(\sigma \xi| \tilde{\xi})$ is the eigenvector for $w_\xi$.
By taking $U(\sigma \xi| \tilde{\xi})$ with $ \tilde{\xi} \in 1,\cdots, M$, we
construct the projection operator $U$, which is a $2M \times M$ rectangular matrix. 
Operating $U$ to the  spin variables on  each ``edge'' of $A$,  we then make the
renormalized local tensor $\tilde{A}$:   
\begin{eqnarray}
\tilde{A} \left\{ \matrix{\tau_{ij} \cr \tilde{\xi}_{ij} \cr } \right\}&\equiv& 
\sum_{ \left\{\sigma_{ij} \right\} \left\{\xi_{ij} \right\} } 
 A\left( \begin{array}{cccc}
\tau_{ij}&\tau_{i'j}&\tau_{i'j'}&\tau_{ij'} \\
\sigma_{ij}&\sigma_{i'j}&\sigma_{i'j'}&\sigma_{ij'} \\
\xi_{ij}&\xi_{i'j}&\xi_{i'j'}&\xi_{ij'}\end{array} \right) 
U( \sigma_{ij} \xi_{ij}|\tilde{\xi}_{ij} ) 
U( \sigma_{i'j} \xi_{i'j}|\tilde{\xi}_{i'j} )
U( \sigma_{i'j'} \xi_{i'j'}|\tilde{\xi}_{i'j'} )  \nonumber \\
&\times&U( \sigma_{ij'} \xi_{ij'}|\tilde{\xi}_{ij'} ) \label{eq:renormtensor}.
\end{eqnarray}

Using Eqs.~(\ref{eq:extension}) and (\ref{eq:renormtensor}) recursively,
we can now calculate the effective local tensor $\tilde{A}_\infty$.
However we encounter another problem in this process; it is also a numerically
heavy problem  to  compute the vertical density matrix with
Eq.~(\ref{eq:getrho}). 

\begin{figure}
\epsfxsize=2.6 in\centerline{\epsffile{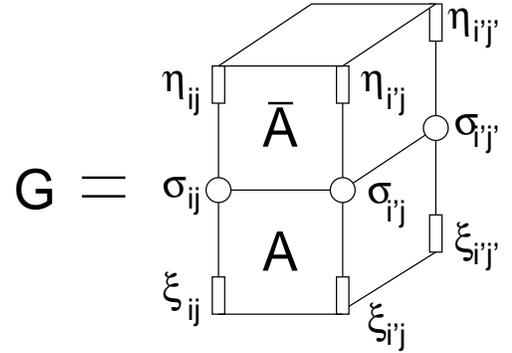}}
\caption{The effective Boltzmann weight $G$.}
\label{fig:newbw}
\end{figure}

Let us next  explain  how to overcome the difficulty in calculating  the
vertical density matrix. The key idea is that we can consider
Eq.~(\ref{eq:getrho}) as  a kind of  2D classical spin system with a point
defect. To see it, we here define a new tensor  
\begin{eqnarray}
G\left\{  \matrix{\eta_{ij}\cr \sigma_{ij}\cr \xi_{ij}\cr}  \right\} \equiv
\bar{A} \left\{ \matrix{ \eta_{ij} \cr \sigma_{ij} \cr} \right\}
A \left\{ \matrix{ \sigma_{ij} \cr \xi_{ij} \cr} \right\}, \label{defg}
\end{eqnarray}
which is graphically represented in Fig.~\ref{fig:newbw}.


Regarding the spin variable along the $z$ axis as a $2M^2$ state single spin: 
\begin{equation}
\bsigma_{ij} = \pmatrix{ \eta_{ij} \cr \sigma_{ij} \cr \xi_{ij} },
\end{equation}
we can see that $G\left\{\bsigma_{ij}\right\}$ becomes  the  Boltzmann weight
for the 2D effective classical model. In fact, the vertical density matrix
$\rho$ can be expressed as the density matrix for the  point defect in the 2D
classical model: 
\begin{eqnarray}
\rho_{kl}(\sigma'_{kl} \xi'_{kl}|\sigma_{kl} \xi_{kl})=
 \sumdash_{[\bsigma]} 
\left[ \proddash_{i+j=even} G \left\{  \bsigma_{ij} \right\} \right] 
G\left\{\check{\bsigma}_{k''l''} \right\}G\left\{\bsigma_{kl} \right\} 
\label{eq:getrho2}
\end{eqnarray}
where the meaning of the prime at the summation and the product is the same as
Eq.~(\ref{eq:getrho}), and the ``checked spin'' is given by  
\[
\left\{ \check{\bsigma}_{k''l''}\right\}  =
\left\{ \matrix{ \eta_{k''l''} \cr \check{\sigma}_{k''l''} \cr \check{\xi}_{k''l''} } \right\} .
\]

We calculate each component of the vertical density matrix~(\ref{eq:getrho2}) as
a statistical average of a special local observable in 2D. To this end we apply
the corner transfer matrix renormalization group (CTMRG)~\cite{CTMRG-ori} which
has been known to be quite efficient for 2D classical statistical systems.

Thus we have obtained a closed algorithm to calculate the local tensor $A$ with
Eqs. (\ref{eq:extension}) and (\ref{eq:renormtensor}), assisted by the CTMRG for
the 2D effective classical model. We summarize the numerical procedure as
follows:
\begin{itemize}

\item[(a)] For the free-boundary condition in the $Z$ direction, define the
initial local tensor 
$A_1$ as
\begin{equation}
A_1\left\{ \matrix{\sigma_{ij} \cr \sigma_{ij}'} \right\}= 
W \left\{ \matrix{\sigma_{ij} \cr \sigma_{ij}'} \right\}.
\end{equation}
For the Ferro boundary condition, define $A_1$ as
\begin{equation}
A_1\left\{ \matrix{\sigma_{ij} \cr \sigma_{ij}'} \right\}= 
W \left\{ \matrix{\sigma_{ij} \cr \sigma_{ij}'} \right\} \times
W \left\{ \matrix{\sigma_{ij}' \cr +1} \right\},
\end{equation}
where $\{+1\}$ means that the four Ising spins are aligned upward.

\item[(b)] Define the effective Boltzmann weight of the 2D classical spin system
with Eq.~(\ref{defg}). 

\item[(c)] Perform the CTMRG calculation for the 2D effective classical spin
system with the Boltzmann weight $G$ and obtain the vertical density matrix
$\rho$. 

\item[(d)] Diagonalize $\rho$ and construct the projection operator $U$.

\item[(e)] Renormalize the local tensor $A$ with Eq.~(\ref{eq:renormtensor}).

\item[(f)] Return (b), until the local tensor $A$ is  converged.

\end{itemize}

In this VDMA calculation, the accuracy is determined  by the number of retained
basis $M$ for the auxiliary variables $\xi$ and $\eta$, and $m$ for the CTMRG
calculation in 2D classical system. We can check the convergence of the computed
quantities with respect to $M$ and $m$.

\section{Results}
\label{sec:result-3dIsing}
\subsection{The 3D Ising model}
\begin{figure}
\epsfxsize=2.6 in\centerline{\epsffile{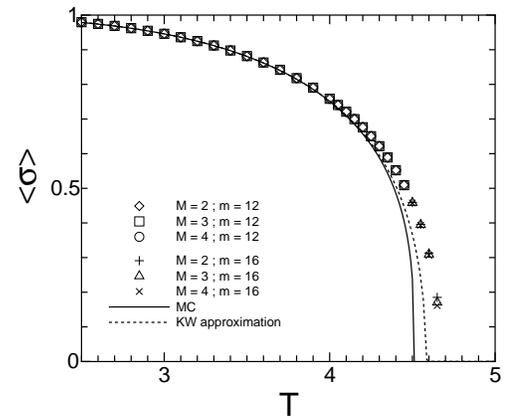}}
\caption{The spontaneous magnetization of the 3D Ising model.}
\label{fig:3dising}
\end{figure}

Fig.~\ref{fig:3dising} shows the spontaneous magnetization $\langle \sigma
\rangle$ calculated by using the VDMA. For comparison we also show  the results
of the 3D-version of the Kramers-Wannier (KW)
approximation~\cite{Okunishi-Nishino-KW} and  the  Talapov and Bl\"ote's Monte
Carlo results~\cite{Trasov}. For each $M$, we have carried out the VDMA
calculations for $m=4$, $8$, $12$, and $16$, where we can find the good
convergence with respect to $m$. We observe that the convergence with respect to
$M$ is also sufficient in the off-critical region. Near the critical point,
however, the magnetization becomes smaller as $M$ is increased, implying that
large $M$ is needed for calculation in the critical region.


\subsection{The 2D transverse field Ising model}
\label{sec:result-transverseIsing}
Let us next consider the VDMA for  the 2D transverse field Ising (TFI) model on
the square-lattice, which is one of the fundamental quantum spin models in 2D.
The Hamiltonian of the 2D TFI model is given by
\begin{equation}
 H = -J\sum_{\langle i,j \rangle} \sigma_i^z \sigma_j^z  - \Gamma \sum_i \sigma_i^x,
\end{equation}
where $J(>0)$ is the ferromagnetic coupling constant,  $\sigma_i^{\alpha}$
($\alpha=x,y,$or $z$) are the Pauli matrices, and $\langle i,j \rangle$ denotes
the nearest-neighbor pairs on the square lattice. 
The transverse field $\Gamma$ induces the quantum fluctuation into the system.
At zero temperature, the TFI model exhibits the quantum phase transition at
$\Gamma = \Gamma_c$, below which the spontaneous magnetization $\langle \sigma
\rangle$ behaves as $\langle\sigma \rangle \sim (\Gamma_c - \Gamma)^{\beta}$.
The critical field $\Gamma_c$ and the critical exponent $\beta$ have been
estimated as $\Gamma_c =3.06$ and $\beta=0.31$ by the quantum Monte Carlo (QMC)
simulation.~\cite{IkeMiy} In the following we consider the VDMA for the 2D TFI
model at zero temperature, and show the results of $\langle \sigma \rangle$.

As was described in the previous section, the VDMA is formulated for the 3D
classical systems. 
In order to apply the VDMA to the 2D TFI model, we map the model to the 3D
anisotropic Ising model by using the Suzuki-Trotter
transformation~\cite{Suzuki}. The partition function $Z$ of the 2D TFI model is
obtained as the limit of the 3D anisotropic Ising model:
\begin{eqnarray}
   Z= \lim_{L \rightarrow \infty}
{\rm Tr} \exp \biggl[ K_h \sum_{\tau=1}^{L}\sum_{\langle i,j
\rangle}\sigma_{i,\tau} \sigma_{j,\tau} 
+ K_v \sum_{\tau=1}^{L}\sum_{i} \sigma_{i,\tau}\sigma_{i,\tau+1} \biggr],
\label{tfip}
\end{eqnarray}
where  $\sigma_{i,\tau}$ is the Ising variable at the position $i$ and imaginary
time $\tau$. The effective couplings $K_h$ and $K_v$ in Eq.~(\ref{tfip}) are
given by 
\begin{eqnarray}
 K_h = \epsilon J, \\
 K_v = -\frac{1}{2}\log[ \tanh(\epsilon \Gamma)],\\
 \epsilon = 1/(T L),
\end{eqnarray}
where the subscripts $h$ and $v$ denote the horizontal ($X$-$Y$) direction and
the vertical (Trotter) one, respectively. We can perform the VDMA calculation
for this anisotropic 3D Ising model. 

We should make a comment about the boundary condition, before proceeding to
details. As was seen in the previous section, the open boundary condition is
assumed in the VDMA. However, the periodic boundary condition is imposed along
the Trotter direction in Eq.~(\ref{tfip}). As far as the zero-temperature
properties are concerned, the boundary condition is inessential due to the
double limit $T\to 0$ and $L\to 0$, allowing us to apply the VDMA to
Eq.~(\ref{tfip}).~\cite{barshas} 

For a fixed value of $\epsilon$, we calculate the magnetization
$\langle \sigma (\epsilon) \rangle$ with the VDMA for the infinite volume. 
After obtaining  $\langle \sigma (\epsilon ) \rangle$ for various $\epsilon$,
we take the $\epsilon \to 0$ limit by extrapolation. In the actual calculation,
we have observed the following $\epsilon$-dependence
\begin{equation}
\langle \sigma (\epsilon ) \rangle = \langle \sigma ( 0 ) \rangle +
{\rm const} \times \epsilon^{2},\label{extrapo}
\end{equation}
which we adopt for the extrapolation formula.

\begin{figure}
\epsfxsize=2.6 in\centerline{\epsffile{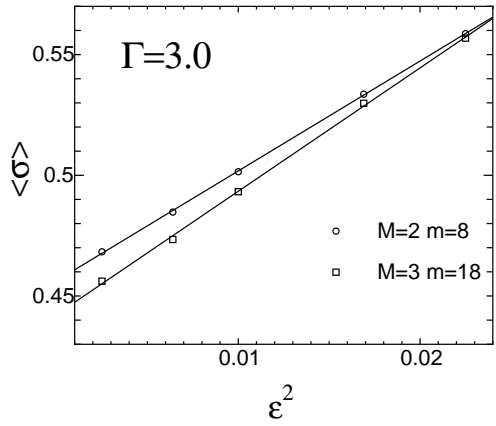}}
\caption{The extrapolation of the magnetization at $\Gamma=3.0$.
The solid lines are linear fits of data.}
\label{gamma30}
\end{figure}

In Fig.~\ref{gamma30}, we show the  $\langle \sigma \rangle$ .vs.  $\epsilon^2$
plots at $\Gamma=3.0$ as an example, which are obtained with the VDMA of the
numbers of retained bases $(M,m)=(2,8)$ and $(3,18)$, where the convergence with
respect to $m$ is rapid. In the region $\Gamma > 3.2$, $\langle \sigma \rangle$
converges to 0 smoothly. 

\begin{figure}
\epsfxsize=2.6 in\centerline{\epsffile{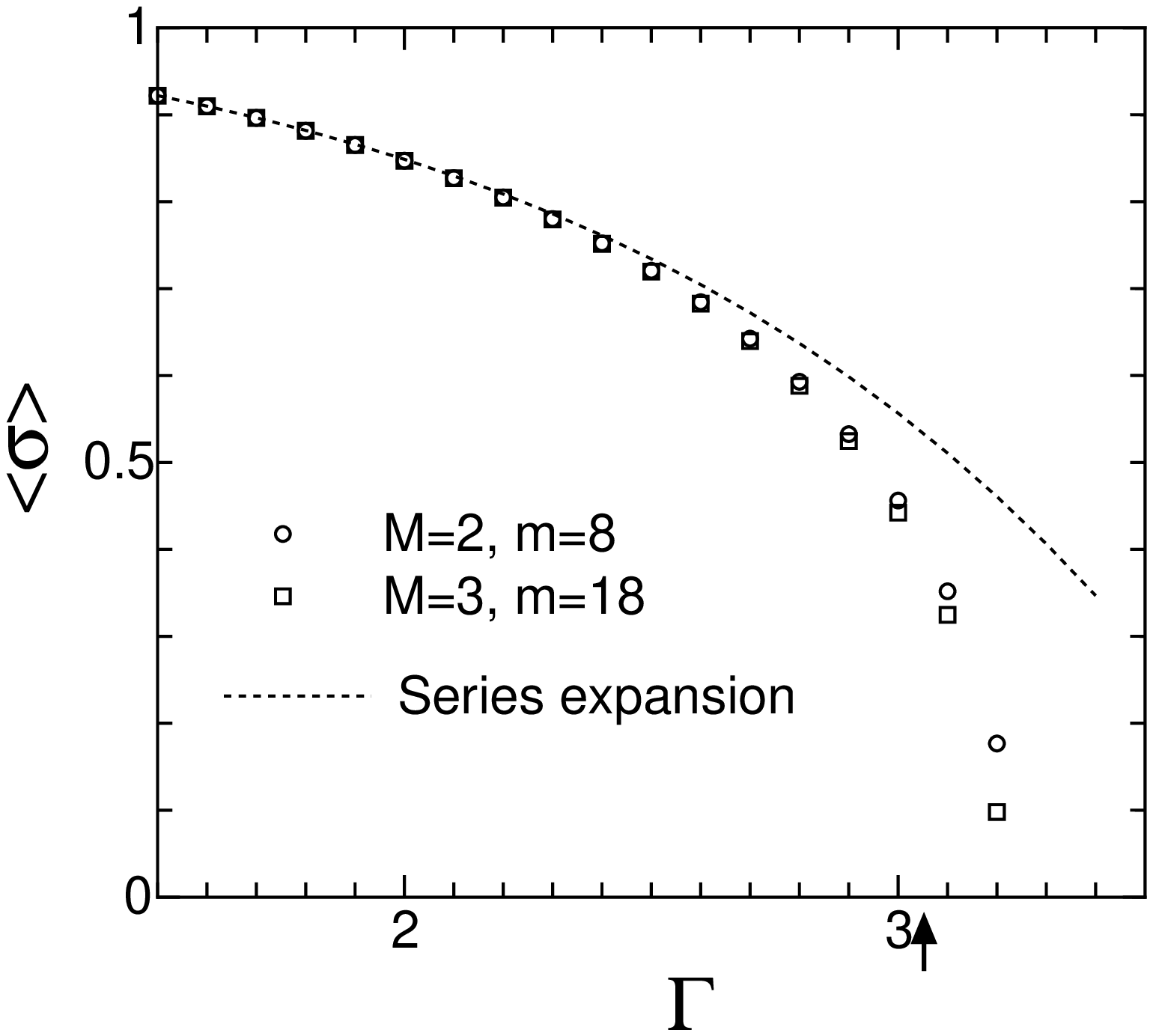}}
\caption{ $\epsilon=0$ limit of $\langle \sigma \rangle$-$\Gamma$ curve at
$T=0$. The arrow shows  
the critical field obtained by the QMC simulation~[17].}
\label{mage000}
\end{figure}

In Fig.~\ref{mage000}, we show the $\langle \sigma \rangle$-$\Gamma$ curve and
the results of the series expansion for comparison\cite{PfeEll}.
In $\Gamma < 2.6$, the VDMA results are sufficiently reliable, where the good
convergence about $m$ and $M$ can be seen.
We can further  see the good agreement with the series expansion in the small
field region ($\Gamma < 2.0$). In the vicinity of the critical point, however,
the calculated magnetization exhibits the $M$ dependence. The roughly estimated
critical field from the VDMA calculation is about 3.2, which is $4\%$ larger
than the QMC one. 


\section{Conclusion}
In this paper we have constructed a novel higher-dimensional numerical
renormalization algorithm  which utilizes the natural tensor-product form of
maximum-eigenvalue eigenstate $|\psi_{\rm max}\rangle$  of the transfer matrix.
In our algorithm, called VDMA (vertical density matrix algorithm), the local
tensor forming $|\psi_{\rm max}\rangle$ is iteratively updated using the
vertical density matrix.  We have successfully applied the VDMA to 3D Ising
model. The VDMA can also be applied to 2D quantum systems using the
Suzuki-Trotter transformation to 3D classical statistical systems, which we have
demonstrated for the 2D  transverse field Ising model.  Application of the VDMA
to other 2D quantum systems such as the Heisenberg model is an important subject
of study, which we are now undertaking.  

\section*{Acknowledgments}
Y. H. and K. O. was  supported by the Japan Society for the Promotion of Science
(JSPS). This work was partially supported by the ``Research for the Future''
program from the JSPS (JSPS-RFTF97P00201) and by the Grant-in-Aid for Scientific
Research from Ministry of Education, Science, Sports and Culture (No. 12640393
and No. 11640376).  


\narrowtext

\widetext

\begin{thebibliography}{99}
\bibitem{BdotB}
In Ref.~\cite{orig-DMRG1} the superblock is constructed as $B \bullet \bullet
B$, where  $B$ and $\bullet$ represent the block and the single site,
respectively. In the VDMA algorithm, however, we construct the superblock as  $B
\bullet B_e$. $B_e$ represents the environment block and $B \ne B_e$. 
In this case, the environment block $B_e$ contains much more sites than the
system block $B \bullet$ that corresponds to $(\sigma_{kl} \xi_{kl} )$.
For example, $B_e$ contains $\eta_{kl}, \eta_{k+1l-1}, \eta_{k-1l+1}, \cdots$
(See Fig.~\ref{fig:vdm}).
\end{thebibliography}
\end{document}